\documentclass[conference]{IEEEtran}
\IEEEoverridecommandlockouts
\usepackage[hyphens]{url}
\usepackage{hyperref}

\usepackage[utf8]{inputenc}
\usepackage{graphicx}
\usepackage{longtable}

\usepackage{ragged2e}
\usepackage{todonotes}
\usepackage[nottoc]{tocbibind}
\usepackage{enumerate}
\usepackage{comment}
\usepackage{threeparttable}

\usepackage{amsmath,amssymb,amsfonts}
\usepackage{algorithmic}
\usepackage{graphicx}
\usepackage{textcomp}
\usepackage{xcolor}
\usepackage{graphicx}   
\usepackage{wrapfig}    
\usepackage{float}      
\usepackage{url}
\usepackage[backend=bibtex,style=ieee,natbib=true]{biblatex} 
\def\BibTeX{{\rm B\kern-.05em{\sc i\kern-.025em b}\kern-.08em
    T\kern-.1667em\lower.7ex\hbox{E}\kern-.125emX}}

\begin{document}

\title{Consensus in Blockchain Systems with Low Network Throughput: A Systematic Mapping Study\\
}

\author{\IEEEauthorblockN{1\textsuperscript{st} Henrik Knudsen}
\IEEEauthorblockA{\textit{Department of Computer Science} \\
\textit{Norwegian University of Science and Technology}\\
Trondheim, Norway \\
henriknu@stud.ntnu.no}
\and
\IEEEauthorblockN{2\textsuperscript{nd} Jakob Svennevik Notland}
\IEEEauthorblockA{\textit{Department of Computer Science} \\
\textit{Norwegian University of Science and Technology}\\
Trondheim, Norway \\
jakob.notland@ntnu.no}
\and
\IEEEauthorblockN{3\textsuperscript{rd} Peter Halland Haro}
\IEEEauthorblockA{\textit{Sintef Nord AS} \\
\textit{Sintef Nord}\\
Tromsø, Norway \\
peter.haro@sintef.no}
\and
\IEEEauthorblockN{4\textsuperscript{th} Truls Bakkejord Ræder}
\IEEEauthorblockA{\textit{Sintef Nord AS} \\
\textit{Sintef Nord}\\
Tromsø, Norway \\
truls.rader@sintef.no}
\and
\IEEEauthorblockN{5\textsuperscript{th} Jingyue Li}
\IEEEauthorblockA{\textit{Department of Computer Science} \\
\textit{Norwegian University of Science and Technology}\\
Trondheim, Norway \\
jingyue.li@ntnu.no}
}

\maketitle

\begin{abstract}
Blockchain technologies originate from cryptocurrencies. Thus, most blockchain technologies assume an environment with a fast and stable network. However, in some blockchain-based systems, e.g., supply chain management (SCM) systems, some Internet of Things (IOT) nodes can only rely on the low-quality network sometimes to achieve consensus. Thus, it is critical to understand the applicability of existing consensus algorithms in such environments. We performed a systematic mapping study to evaluate and compare existing consensus mechanisms' capability to provide integrity and security with varying network properties. Our study identified 25 state-of-the-art consensus algorithms from published and preprint literature. We categorized and compared the consensus algorithms qualitatively based on established performance and integrity metrics and well-known blockchain security issues. Results show that consensus algorithms rely on the synchronous network for correctness cannot provide the expected integrity. Such consensus algorithms may also be vulnerable to distributed-denial-of-service (DDOS) and routing attacks, given limited network throughput. Conversely, asynchronous consensus algorithms, e.g., Honey-BadgerBFT, are deemed more robust against many of these attacks and may provide high integrity in asynchrony events.

\end{abstract}

\section{Introduction}
Blockchain has evolved significantly since its initial roots from Bitcoin and has seen adoption with novel use of the technology for healthcare \cite{mettler2016blockchain}, banking \cite{peters2016understanding}, control systems \cite{stanciu2017blockchain}, and SCM \cite{chang2020blockchain}. In SCM systems, different actors are working together to deliver timely and quality products to their customers. A consensus algorithm used in SCM systems shall provide high transaction throughput to take advantage of vast amounts of IOT sensor data. Furthermore, it would need to function well in an environment with varying network delays. Some nodes, e.g., nodes to collect and transfer temperature data of fresh food on trucks or vessels, may have poor or no internet connection. The blockchain system will be subject to the CAP theorem \citep{CAP}, which is challenging, given high requirements towards transaction throughput and the constraint caused by an unstable network environment. 

This study aims to study to which degree existing consensus algorithms can provide integrity and security with limited network throughput and low network quality. Our primary research hypothesis theorizes that there is little empirical knowledge of blockchain applications' effectiveness with limited network throughput concerning the consensus mechanisms. We performed a mapping study and covered published and preprint articles. Our study aimed at answering the following research questions.

\textbf{RQ1:} How well existing consensus algorithms can provide integrity in an environment with limited network throughput?

\textbf{RQ2:} How well existing consensus algorithms can provide security in an environment with limited network throughput?

We identified and analyzed 25 consensus algorithms qualitatively in this study. The results show that many existing consensus algorithms are unfit for use in an environment affected by varying network throughput. Some consensus algorithms which assume a partially synchronous network can provide integrity in events of asynchrony. However, their transaction throughput may be significantly reduced when faced with targeted denial-of-service attacks. Consensus algorithms that do not make synchronization assumptions, adapted to an asynchronous network, may avoid both.  

The remainder of the paper is organized as follows. 
Section 2 lists related work. 
Section 3 explains the research design and implementation. 
Section 4 present the research result.
Section 5 discusses the results and Section 6 concludes the study.

\section{Related Work}
\label{sec:related_work}

Studies, e.g., \citep{RN21, RN18, RN19, RN22, RN25}, have focused on reviewing the state of the art consensus algorithms.
\citep{RN21} defined a five-component framework for categorizing blockchain consensus algorithms, provided a comprehensive review of current consensus algorithms, and evaluated the algorithms' performance according to their fault tolerance and throughput. \citep{RN18} analyzed the algorithm's throughput, mining incentive, decentralization, and security challenges. \citep{RN25} provided a game-theoretic point of view and looked at the mining incentive provided by different consensus algorithms. \citet{RN19} evaluated the Proof-of-Work (PoW) scheme of cryptocurrencies like Bitcoin and Ethereum according to traditional Byzantine consensus algorithms, and provided insight into PoW specific security challenges, like the Bitcoin anomaly and balancing attacks. \citep{RN22} provided an extensive mapping of consensus algorithms, according to their architecture and paradigm, and highlighted fundamental differences between public and private blockchain systems, and showed how this impacts the algorithms' applicability in these specific types of systems. 

Other studies, e.g., \citep{RN17, RN30}, have focused on security challenges related to using using blockchain technology. In particular, the studies highlighted the security implications the choice of consensus algorithm has upon the overall blockchain system. \citep{RN17} provided a comprehensive overview of security and privacy aspects of blockchain technology, defined core security properties, and described existing security techniques. \citet{RN30} assessed major Nakamoto style consensus algorithms against the 51\% attack, as well as other major security threats towards blockchain systems, and reviewed current mitigation techniques. 

Both industry and academia have shown great interest in assessing new use cases for blockchain technology, following its success within the cryptocurrency space \citep{RN24, RN16, RN23, RN26}. \citet{RN24} provided a \textit{vademecum} to guide designers in their decisions for when and how to apply blockchain technology to their specific use case. Reference \citep{RN16} presented a systematic review of blockchain within the energy sector and discussed its limitations and potential use cases. Reference \citep{RN23} reviewed current blockchain initiatives and highlighted domains having real-world technology adoption. \citet{RN26} presented a comprehensive analysis of the state of the art consensus algorithms and their appropriateness related to cyber-physical systems. In particular, the authors outlined domain-specific challenges for supply chain management blockchain systems.

Another exciting research area has been the interconnection of blockchain technology and IOT systems. Several studies, e.g., \citep{RN27, RN31}, have been dedicated to describing the state of the art IOT blockchain systems and their challenges. Reference \citep{RN27} provided an extensive analysis of key components of IOT blockchain systems and promising consensus schemes and reviewed major IOT blockchain projects. \citep{RN31} reviewed core security issues related to IOT systems and how blockchain technology might be applied to solve some of these issues. The research goals of the related studies mentioned above are summarized in \autoref{tab:relevant_studies_characteristics}.  Furthermore, the consensus algorithms they discussed are summarized in \autoref{tab:relevant_studies_consensus_algos}.
\renewcommand{\arraystretch}{1.25}

\begin{table*}[t]
    \centering
     \begin{tabular}{p{6cm}ccccccccccccc}
        Reference ID & \cite{RN21} & \citep{RN18} & \cite{RN19} & \cite{RN22} & \cite{RN25} & \cite{RN17} & \cite{RN30} & \cite{RN24} & \cite{RN16} & \cite{RN23} & \cite{RN26} & \cite{RN27} & \cite{RN31} \\\hline
        Review existing consensus algorithms & x & x & x & x & x & x & x & x & x & x & x & x & x \\
        Review and analyze current blockchain projects &  &  & x &  &  &  &  & x & x & x &  & x & x \\
        Security analysis &  & x & x &  & & x & x &  &  &  &  &  &  \\
        Provide analysis framework & x & x &  &  &  &  &  &  &  &  &  &  &  \\
        Present domain specific use cases &  & &  &  &  &  &  & x & x & x & x & x & x \\
    \end{tabular}
     \caption{Research goals of related studies }
     \label{tab:relevant_studies_characteristics}
\end{table*}

\begin{table*}[t]
    \centering
    \begin{tabular}{p{6cm}ccccccccccccc}
        Reference ID & \cite{RN21} & \citep{RN18} & \cite{RN19} & \cite{RN22} & \cite{RN25} & \cite{RN17} & \cite{RN30} & \cite{RN24} & \cite{RN16} & \cite{RN23} & \cite{RN26} & \cite{RN27} & \cite{RN31} \\\hline
    
     Proof of Work (PoW) & x & x & x & x & x & x & x & x & x & x & x & x & x \\
     Proof of Stake (PoS) & x & x &  & x & x & x & x & x & x & x & x & x & x \\
     Delegated Proof of Stake (DPOS) & x & x &  & x &  & x & x & x & x & x & x &  & x \\
     Proof of Authority (PoAuth) & x &  &  & x &  & x &  & x & x & x &  & x &  \\
     Proof of Elapsed Time (PoET) & x & x &  & x & x & x &  & x & x &  & x &  &  \\
     Proof of TEE-Stake (PoTS) & x &  &  &  &  &  &  &  &  &  & x &  &  \\
     Proof of Retrievability (PoR) & x &  &  & x & x &  &  &  &  &  &  &  &  \\
     Proof of Weight (PoWeight) &  & x &  &  &  &  &  &  &  &  &  &  &  \\
     Proof of Burn (PoB) &  & x &  & x & x &  &  &  & x & x & x & x & x \\
     Proof of Capacity (PoC) &  & x &  & x &  &  &  &  & x & x & x & x & x \\
     Proof of Importance (PoI) &  & x &  & x &  &  &  & x &  & x & x & x &  \\
     Proof of Authority (PoA) &  & x &  & x & x &  &  &  & x & x &  &  &  \\
     Practical Byzantine Fault Tolerance (PBFT) & x & x & x & x &  & x &  & x & x & x & x & x & x \\
     Delegated Byzantine Fault Tolerance (Delegated BFT) &  & x &  & x &  &  &  &  &  &  & x &  &  \\
     Democratic Byzantine Fault Tolerance (Democratic BFT) &  &  & x &  &  &  &  &  &  &  &  &  &  \\
     Byzantine Fault Tolerant State Machine Replication (BFT-SMART) &  &  &  & x &  &  &  &  &  &  &  &  &  \\
     Honey-Badger Byzantine Fault Tolerance (Honey-Badger BFT) & x &  &  & x &  & x &  &  &  &  &  &  &  \\
     Ripple Protocol Consensus Algorithm (RPCA) & x &  & x & x &  &  &  & x &  & x & x & x &  \\
     Stellar Consensus Protocol (SCP) &  &  &  & x &  &  &  & x &  &  &  &  &  \\
     Byzantine Fault Tolerance based Proof of Work (BFT-based POW) &  &  &  &  & x &  &  & x &  &  &  & &  \\
     Byzantine Fault Folerance based Proof of Stake (BFT-based POS) & x &  &  & x & x & x &  & x &  &  &  &  & \\
     Paxos &  &  &  &  &  &  &  &  &  &  & x &  &  \\
     Raft &  &  &  & x &  & x &  & x &  &  & x &  &  \\
    \end{tabular}
    \caption{Consensus algorithms  in related studies}
     \label{tab:relevant_studies_consensus_algos}
\end{table*}



None of the related studies highlight the challenge of consensus in environments with varying network delays. In particular, in the context of supply chain management systems, real-world scenarios often involve entities transporting goods over larger geographical areas, with periods of poor or no internet connection. To deliver on the promises of traceability, accountability, and improved information sharing, SCM blockchain systems' underlying consensus algorithm must allow for efficient data sharing between stakeholders, despite participants being affected by varying network delays. Therefore, assessing the state of the art consensus algorithms' performance under such an environment is essential to find suitable consensus algorithms for SCM blockchain systems. Furthermore, such an analysis may uncover the need for designing novel consensus mechanisms if current mechanisms are found insufficient for the use case.

\section{Research Design and Implementation}

This study aims to provide insight into core mechanisms of the state of the art consensus algorithms and how they relate to SCM blockchain systems in an enterprise setting with multiple stakeholders affected by varying network throughput. 

\subsection{Data collection}
We followed the systematic mapping study guideline proposed by \citep{RN14}. The search query \textit{"Blockchain AND Consensus Algorithm"} was used to identify papers related to consensus in blockchain systems. The query was executed at Oria \citep{Oria}, a search engine aggregating research papers from scientific databases, including IEEE Xplore, Springer, ACM Digital library, and Scopus. We included only peer-reviewed research papers in this round of search. Furthermore, to also gain insight into the current non-peer-reviewed literature, the aforementioned search query was also executed towards Arxiv.org. This combined search is meant to provide a holistic view of the current literature on consensus mechanisms applicable to blockchain systems.
To exclude irrelevant papers, we first read through the papers' titles and abstracts. Papers that did not explicitly mention consensus algorithms were excluded. Papers that were not available online were also excluded. Furthermore, we excluded any paper that does not provide an adequate description of the algorithm or sufficient security analysis. 

A total of 2151 papers were identified via the Oria search engine. Through disregarding non-peer-reviewed papers, the total was narrowed down to 451 papers. We then read through the papers' titles and abstracts to exclude any paper that did not explicitly mention consensus algorithms and got 81 papers. Of these 81 papers, 63 papers were identified as studies that presented consensus algorithms.  Regarding the search executed at  Arxiv.org,  a  total of  103  papers were identified.  The papers were analyzed in the same manner as with the  Oria search. To get representative consensus algorithms, We only include consensus algorithms if they were discussed in more than one paper. We end up with a total of 25 consensus algorithms to analyze.



 

\subsection{Data analysis}


Our classification is based on qualitative analysis and uses a classification scheme differing from \cite{RN14} because none of the identified papers sufficiently covered the topic of consensus in low throughput networks. To classify performance of the consensus algorithms and to answer RQ1, we applied the metrics shown in \autoref{tab:classification_scheme}. 

\begin{itemize}

\item Reference \citep{RN52} define read and transaction latency and read and transaction throughput as essential performance metrics. In this study, we are specifically interested in evaluating consensus mechanisms in relation to a blockchain system affected by low network throughput, distributed over a greater geographical area. Thus, the particular focus of the classification is attributed to communication-related costs, in the form of \textit{consensus latency} and \textit{communication complexity}.

\item To avoid the impossibility result \citep{RN40}, one can assume that the underlying communication network is synchronous. However, the network may be particular asynchronous \citep{RN46} because of the limited network throughput. Thus, the consensus algorithms' \textit{timing assumptions} are also included as a metric. The timing assumptions made by consensus algorithms can be categorized accordingly. 
\begin{itemize}
    \item \textbf{Synchronous:} There exist a known fixed upper bound $\Delta$ on the message delay between peers within the network. 
    \item \textbf{Partial Synchronous:} Either of the following statements holds: 
    
      \begin{enumerate}[i]
      \item There exists an upper bound $\Delta$ on the message delay between peers within the network, but it cannot be known a priori.
      \item There exists an upper bound $\Delta$ on the message delay between peers within the network, but it does not hold before an unknown point of time $T$.
      \end{enumerate}
   \item \textbf{Asynchronous}: There is no known fixed upper bound $\Delta$ on the message delay between peers within the network.
\end{itemize}

\item \textit{Byzantine fault tolerance} and transaction \textit{finality} are included, highlighting the algorithms' resilience towards adversarial attacks, as well as transaction confirmation time. A consensus algorithm's transaction finality is the algorithm's guarantee that committed transactions cannot be reversed. Some consensus algorithms providing probabilistic finality, e.g., Nakamoto style consensus, favor availability over strong consistency. Other consensus algorithms providing immediate finality, e.g., BFT style consensus, need strong consistency to enforce this, thus sacrificing the system's availability during network partitioning.  

\end{itemize}

\begin{table}[h]
    \centering
     \begin{tabular}{lp{5cm}}
        Metrics & Description \\\hline
        Consensus latency &  Number of Round-trip time (RTT) needed to complete a round of consensus.  \\
        Communication complexity &  Number of messages needed to complete a round of consensus.  \\
        Timing assumptions &  Timing assumptions made by the consensus model related to the underlying network. This relates to a synchronous, partial synchronous or asynchronous network.  \\
        Byzantine Fault Tolerance & The percentage of adversarial control in the network, in which the consensus model can resist double spend attacks.  \\
        Finality & The assurance that transactions committed will not be reverted. Either immediate or probabilistic.  \\
        
    \end{tabular}
    \caption{Metrics to measure performance and to answer RQ1}
    \label{tab:classification_scheme}
  
\end{table}

        
  





To answer RQ2, we first identified core security issues related to the blockchain system based on \cite{RN17} and \cite{RN30}. The identified security issues are shown in \autoref{tab:background_security_considerations}. In terms of the SCM setting, with a permissioned blockchain system affected by limited network throughput, not all security issues listed in \autoref{tab:background_security_considerations} are relevant. For example, given that participants in the system are authenticated, a Sybil attack can effectively be combated. In an authenticated setting, aspects like user anonymity and transaction unlinkability are deemed irrelevant. We excluded irrelevant security issues and classified the consensus algorithms based only on double-spend attacks, balance attacks, Long-range attacks, p + Epsilon attacks, DDOS attacks, and BGP attacks.

\begin{table*}[!ht]
    \centering
    \begin{tabular}{p{4cm}p{13cm}}
         Name & Description \\\hline
         Consistency & Which approach will the system utilize to ensure the consistency of the system's ledger?  \\
         Tamper-Resistance & Is the system able to ensure the integrity of the ledger? \\
         Byzantine Fault Tolerance & To which an extent do the system need to tolerate Byzantine faults, e.g. adversarial nodes?  \\
         Sybil attack & How will the system protect against malicious actors creating multiple fake identities, attempting to outvote an honest majority?  \\
         Double spend attack & How will the system prevent nodes spending the same currency for two separate transactions? Specifically related to cryptocurrency systems.  \\
         Long-Range attack & To which extent is the system vulnerable to Long-Range attacks; forking the blockchain at its genesis block and privately building an alternative chain?  \\
         P+ Epsilon attack & To which extent is the system vulnerable to P+ Epsilon attacks; taking advantage of the dominant strategy among network participants, leveraging non-altruistic participants against the system? \\
         Balance attack & To which an extent is the system vulnerable to balancing attacks; splitting the network into sub-networks, delaying transactions and performing double spend attacks? \\
         Border Gateways Protocol (BGP) Hijacking & How will the system handle adversarial attacks against the network's routing mechanisms; partitioning the network or delaying block propagation? \\
         Distributed-Denial-of-Service (DDOS) susceptibility &  To which an extent is the system susceptible to DDOS attacks, in terms of impact upon safety and liveness? Especially relevant for leader based consensus models. \\ 
         Degree of centralization & To which an extent is the system centralized around a subset of participants? This may have implications on various double spend attacks, as well as DDOS susceptibility.   \\
         User anonymity & Are participants of the system able to partake in transactions while staying anonymous?\\
         Transaction confidentiality & To which extent are transactions kept confidential? \\
         Transaction unlinkability & To which extent is it possible to link a group of transactions to a specific identity?\\
    \end{tabular}
    \caption{Security considerations for blockchain systems}
    \label{tab:background_security_considerations}
\end{table*}

\section{Research Results}
\label{sec:study_results}
We first classified the 25 consensus algorithms identified from the literature into five categories, namely, Nakamato, Byzantine Fault Tolerant (BFT), Federated Byzantine Agreement (FBA), Hybrid, and Crash Tolerant, based on their core principles. Then, we analyzed 1) whether the algorithm provides sufficient performance in this specific application scenario with limited network throughput; 2) whether the algorithms provide integrity in the application scenario, as of RQ1; 3) to what extent the algorithms provides security in this application scenario, as of RQ2. The findings related to the performance metrics are summarized in Table \ref{tab:app_classification}. The answers to RQ1 and RQ2 are summarized in Table \ref{tab:discussion_summary}.

\subsection{Performance}

\textbf{Nakamoto style} consensus algorithms usually provide consensus in a single communication step, except PoA. They all operate with linear complexity as well as requiring an honest majority. The timing assumption is synchronous, which is challenging with low network throughput. With probabilistic finality, there is also a great chance of forking given our SCM application scenario.

\textbf{BFT style} consensus algorithms usually provide consensus in a three communication step process, with quadratic communication complexity. Delegated BFT, Mixed BFT, and Linear-BFT can achieve linear communication complexity through partitioning the network. Delegated BFT and Mixed BFT delegate consensus to a subset of the nodes in the network. Linear-BFT reduces the number of messages sent per node by using an expander graph. This, however, comes at the cost of either significantly increased centralization or potentially higher consensus latency. BFT style consensus algorithms all provide deterministic finality, which may be of major benefit, providing low confirmation latency for transactions. In the enterprise setting, this can be a deciding factor. These consensus algorithms can typically withstand a maximum of $\frac{1}{3}$ Byzantine nodes, posing a significantly weaker resistance than the honest majority threshold provided by Nakamoto style consensus algorithms. Linear-BFT emerges as a promising innovation, providing the optimal threshold of $\frac{1}{2}$ Byzantine nodes, while also having amortized optimal communication complexity of $\Theta(n)$. 

\textbf{FBA style} consensus algorithms do not provide any guarantees to the number of consensus steps involved in agreeing upon a set of transactions. As nodes only communicate within their respective quorum slice, communication complexity depends on each node's quorum slice's size. In Ripple, Unique Node Lists (UNL) need to overlap 90\% across the network to ensure security, making communication complexity $\Theta(n^2)$. In SCP, nodes are free to pick their quorum slices based on their own reasoning (e.g., reputation, wealth, brand). In an optimistic setting, where the size of quorum slices is constant, SCP's communication complexity is $\Theta(n)$. SCP furthermore provides optimal resilience towards Byzantine nodes in the federated setting, only requiring that nodes' quorum slices intersect honestly. RPCA makes an assumption of a maximum of 20\% Byzantine nodes for a given UNL list. As FBA agreement builds upon nodes' quorum slices intersecting, consensus relies on that not all these interconnections are faulty. If these critical nodes suffer from low network throughput, this could hamper the speed and throughput in which the network can process transactions. In particular, if they cannot respond for a long time, transaction throughput may halt.

\textbf{Hybrid style} consensus algorithms typically provide consensus in linear communication complexity using a three-step BFT style process to reach consensus, while using mechanisms from Nakamoto style consensus algorithms for leader election. This scheme allows for immediate finality, which enables consensus algorithms to work in a permissionless setting. The consensus algorithms can scale better than standard Nakamoto consensus algorithms with the cost of reduced Byzantine fault tolerance.

\textbf{Crash tolerant style} consensus algorithms, i.e., Raft, provide consensus in a single communication step, with the optimal communication complexity of $\Theta(n)$. It also provides deterministic finality. The main caveat related to crash tolerant consensus algorithms is that they cannot tolerate Byzantine faults. Disregarding Byzantine faults, Raft tolerates up to 50\% crashed nodes. Raft's communication pattern is completely leader-centered. During normal operation, the leader continuously broadcasts messages to the rest of the network, while follower nodes only respond to the messages they receive. Therefore, the leader node must have sufficient network throughput and computational resources for the leader not to become a bottleneck in terms of performance.

\begin{table*}[t]
 \centering
\begin{tabular}{p{3cm}llllrl}
    Name & Ref & Latency & Complexity & Timing assumptions & BFT & Finality \\\hline
     \multicolumn{7}{l}{\textbf{Nakamoto style consensus algorithms}} \\\hline
     PoW & \citep{Nakamoto_Bitcoin_POW} & 1 RTT & $\Theta(n)$ & Synchronous & 50 \% & Probabilistic\\
     PoS & \citep{RN64} & 1 RTT & $\Theta(n)$ & Synchronous & 50 \% & Probabilistic\\
     DPoS & \citep{DPOS_Larimer} & 1 RTT & $\Theta(n)$ & Synchronous   & 50 \%& Probabilistic \\
     DDPoS & \citep{DDPOS_White_IEEE} & 3 RTT & $\Theta(n^2)$ & Synchronous &  50 \% & Probabilistic\\
     PoAuth & \citep{RN66} & 1 RTT & $\Theta(n)$ & Synchronous & 50 \%& Probabilistic \\
     PoET & \citep{POET_SPEC} & 1 RTT & $\Theta(n)$ & Synchronous & 50 \% & Probabilistic\\
     PoTS & \citep{RN67} &  1 RTT & $\Theta(n)$ &  Synchronous& 50 \% & Probabilistic\\
     PoR & \citep{RN68} & 1 RTT & $\Theta(n)$ & Synchronous & 50 \% & Probabilistic\\
     PoB & \citep{RN69} & 1 RTT & $\Theta(n)$ & Synchronous & 50 \% & Probabilistic \\
     PoC & \citep{RN70} & 1 RTT & $\Theta(n)$ &  Synchronous & 50 \% & Probabilistic\\
     PoI & \citep{NEM_POI} & 1 RTT & $\Theta(n)$ &  Synchronous & 50 \% & Probabilistic\\
     PoA &  \citep{POA} & 3 RTT & $\Theta(n)$ &  Synchronous & 50 \% & Probabilistic \\\hline
     \multicolumn{7}{l}{\textbf{BFT style consensus algorithms}} \\\hline
     Linear-BFT & \citep{RN36} & 6 RTT & $\Theta(n)$ & Synchronous & 50 \% & Immediate \\
     PBFT & \citep{PBFT_MIT} & 3 RTT & $\Theta(n^2)$ &  Partial Synchronous & 33 \% & Immediate\\
     IBFT & \citep{RN37} & 3 RTT & $\Theta(n^2)$ & Partial Synchronous & 33 \% & Immediate\\
     Delegated BFT & \citep{RN71} & 3 RTT & $\Theta(n)$ & Partial Synchronous  & 33 \%  & Immediate\\ 
     BFT - SMART &  \citep{BFT_SMART} & 3 RTT & $\Theta(n^2)$ & Partial Synchronous  & 33  \% & Immediate \\
    T-PBFT & \citep{RN34} & 3 RTT & $\Theta(n^2)$ & Partial Synchronous  & 33 \% & Immediate\\ 
      MBFT & \citep{RN33} & N/A & $\Theta(n)$ & Partial Synchronous & 33 \% & Immediate\\
    Honey-BadgerBFT & \citep{RN42} & 6 RTT & $\Theta(n^2)$ & Asynchronous & 33 \% & Immediate\\\hline
    
      \multicolumn{7}{l}{\textbf{Federated Byzantine Agreement consensus algorithms}} \\\hline
     RPCA & \citep{RIPPLE_ANALYSIS} & N/A & $\Theta(n^2)$ &  Partial Synchronous  & 20 \% & Immediate \\
     SCP & \citep{SCP} & N/A & $\Theta(n)$ & Partial Synchronous & N/A & Immediate \\\hline
     \multicolumn{7}{l}{\textbf{Hybrid consensus algorithms}} \\\hline
     BFT-based POW & \citep{RN73} & 3 RTT & $\Theta(n)$ &  Partial Synchronous & 33 \% & Immediate \\ 
     BFT-based POS & \citep{RN72} & 3 RTT & $\Theta(n)$ &  Partial Synchronous & 33 \%  & Immediate\\\hline 
     \multicolumn{7}{l}{\textbf{Crash Tolerant consensus algorithms}} \\\hline
     Raft & \citep{RAFT} & 1 RTT & $\Theta(n)$ &  Partial Synchronous & 0 \% & Immediate \\
\end{tabular}
\caption{Consensus algorithm classification}
    \label{tab:app_classification}
\end{table*}

\begin{table*}[h]
    \centering
    \begin{threeparttable}
    \begin{tabular}{p{4cm}lllll}
    
         \multicolumn{6}{p{14cm}}{\rule{0pt}{6ex}\textbf{RQ1: What consensus models are able to provide integrity in an environment with limited network throughput?}}\\\hline
         \textbf{Consensus algorithms} & Nakamoto & BFT & FBA & Hybrid & Crash Tolerant \\\hline
         Integrity in asynchronous system & - & ! & + & + & + \\
         \multicolumn{6}{p{14cm}}{\rule{0pt}{6ex}\textbf{RQ2: What consensus models are able to provide security in an environment with limited network throughput?}}\\\hline
         \textbf{Consensus algorithms} & Nakamoto & BFT & FBA & Hybrid & Crash Tolerant \\\hline
         Double spend attacks & - & + & + & + & ! \\
         Balance attacks & - & + & + & + & + \\
         Long-Range attacks & ! & + & + & + & + \\
         P + Epsilon attacks & ! & + & + & + & + \\
         DDOS attacks & + & - & ! & - & - \\
         BGP attacks & - & - & ! & - & - \\
    \end{tabular}
    \bigskip
    \begin{flushleft}
        \begin{tablenotes}
            \item[+] Strong guarantees
            \item[-] Lacking guarantees
            \item[!] Depending on the specifics of the consensus algorithm or the configuration of the blockchain system
        \end{tablenotes}
    \end{flushleft}
    \caption{Summary of consensus algorithms in relation to performance metrics, RQ1 and RQ2. }
    \label{tab:discussion_summary}
    \end{threeparttable}
\end{table*}

\subsection{Integrity}
Limited network throughput may cause increased asynchrony within a blockchain system. This makes it infeasible to make assumptions regarding the ordering or timing of messages being sent. As such, consensus algorithms relying on synchronous network for correctness cannot provide integrity in the relevant setting.

\textbf{Nakamoto style} consensus algorithms are unfit for the setting \citep{RN45}. 

\textbf{BFT style} consensus algorithms may work depending on their timing assumptions. Linear-BFT specifically assumes synchronous network and therefore cannot guarantee integrity either. BFT consensus algorithms assuming a partial synchronous or asynchronous network will still provide integrity in such an asynchronous network. It is important to point out that while many BFT style consensus algorithms can provide integrity during times of asynchrony, these same algorithms cannot provide liveness at the same time, as a result of the FLP impossibility \citep{RN40}. The network needs to stabilize before transactions can be continued to be processed. In light of this, the Honey-BadgerBFT consensus algorithm differentiates itself, as it makes no assumptions regarding the network's synchrony. As Honey-BadgerBFT functions purely asynchronously, it may avoid the standstill of asynchrony or performance impact of a larger message delay $\Delta$. In particular, \citep{RN42} points out that there will always be an inherent trade-off related to such synchrony assumptions. If the $\Delta$ parameter value is too low, the system will not provide progress. If the $\Delta$ parameter value is too high, it will not fully take advantage of the network's bandwidth. The authors propose Honey-BadgerBFT as a candidate for consortium blockchain systems, emphasizing its benefits of robustness and high transaction throughput, despite being an asynchronous system.

\textbf{FBA style} consensus algorithms, i.e., RPCA and SCP, assume a partially synchronous network and can provide integrity even in periods of asynchrony. However, they require messages to be bounded by some upper bound message delay $\Delta$ to provide forward progress. 

\textbf{Hybrid style} consensus algorithms Byzcoin and Tendermint make synchronization assumptions equal to that of partial synchrony. In times of asynchrony, consensus algorithms can preserve integrity. Conversely, the consensus algorithms' transaction throughput may suffer significantly during these events, similar to that of BFT style consensus algorithms.

\textbf{Crash tolerant style} consensus algorithm relies on synchronization assumptions similar to that of BFT style consensus algorithms. In particular, Raft assumes a known broadcasting latency, which is then used to determine an election timeout interval. The algorithm can ensure integrity, even in periods of asynchrony, despite the message delay exceeding the assumed bound. However, it will not be able to provide any forward progress in such events. There will be a continuous re-election process until the network stabilizes. This may be problematic in terms of a network with high communication latency and poor network throughput. 

\subsection{Security}

\textbf{Nakamoto style} consensus algorithms are targets of a multitude of attacks. Sufficient delay of messages caused by the asynchrony of the network may enable double spend and balance attacks. Furthermore, the Nakamoto style consensus algorithms may be vulnerable to P + Epsilon attacks, depending on the blockchain system's incentivization mechanisms. In contrast to many other types of consensus schemes, which are centered around a stable leader, Nakamoto style consensus algorithms elect leaders non-deterministically on a block-to-block basis, which makes DDOS attacks infeasible in most practical settings. Conversely, the consensus algorithms may, however, be vulnerable to BGP attacks. In particular, \citep{RN53} has shown the feasibility of utilizing routing attacks against Bitcoin's PoW scheme.

\textbf{BFT style} consensus algorithms usually have an increased susceptibility towards DDOS attacks, compared to Nakamoto style consensus. Specifically, when sufficiently powerful adversaries have insight into which of the network's nodes is the leader for a given election term or view, they could impair the system's liveness by overloading the leader with incoming traffic. This is exasperated in our SCM setting with low network throughput. As many BFT style consensus algorithms rely on a stable leader for progress, in the worst case, this could lead to a continuous view change, grinding transaction throughput to a halt. The Honey-BadgerBFT consensus algorithm once again differentiates itself from the other BFT style consensus algorithms in that it is leaderless. Therefore, the Honey-BadgerBFT consensus algorithm does not suffer from the aforementioned susceptibility towards DDOS attacks. DDOS attacks could also potentially target non-leader nodes in an attempt to increase the number of faulty nodes. As most BFT style consensus algorithms require a supermajority of honest nodes to ensure the system's integrity, a sufficiently powerful adversary could target honest nodes and break the Byzantine Fault Tolerance threshold. Whether this would be feasible in a practical setting depends on the size of the network, the current number of Byzantine nodes, as well as the adversaries' combined computational resources. Linear-BFT, which only requires a majority of honest nodes to ensure integrity, may be better suited to defend against this kind of attack. BFT style consensus algorithms could also be vulnerable to BGP attacks. When an adversary gains control of routing mechanisms used within the system, they could potentially partition the network or delay communication between nodes. Targeting the leader could hamper the liveness of the system. When the adversaries have access to an especially central router within the communication system, they could isolate larger groups of participants. Like a DDOS attack, this could allow adversaries to break the Byzantine Fault Tolerance threshold, invalidating any guarantees about the system's integrity.  

\textbf{FBA style} consensus algorithms avoid some of the pitfalls related to BFT style consensus. In particular, as RPCA and SCP are leaderless, there is no vulnerability related to a leader being targeted by a DDOS attack, stopping the system from achieving progress. There is, however, a new potential attack vector related to splitting the network. Going back to the important role in which intersecting nodes play in federated Byzantine agreement, a sufficiently powerful adversary could perform DDOS attacks targeting well-connected nodes. Such an attack could partition the network, isolating nodes, making consensus in the original network unachievable, and hindering transaction throughput. FBA style consensus algorithms could be vulnerable to the BGP attack. Similar to that of a targeted DDOS attack against a well-connected node, if an adversary gains control of a central router within the communication network, they could partition the network, creating divergence. In terms of RPCA, where the 90\% UNL overlap makes divergence attacks unlikely, attackers might rather try to break the Byzantine Fault Tolerance threshold of 20\%. As for SCP, allowing users to communicate only with nodes selectively they specifically trust, any form of substantial routing attack may break the network, depending on the size of each user's node list.

\textbf{Hybrid style} consensus algorithms have an increased susceptibility to DDOS attacks due to the transition from Nakamoto style consensus to a hybrid approach. Like BFT style consensus algorithms, the hybrid approach is leader-centered and needs to maintain a stable leader to provide high performance. Combined with limited network throughput, this may further hinder transaction throughput. DDOS attacks might also target non-leader nodes of the network. As inclusion in the inner consensus group is the core of Nakamoto style consensus mechanisms, attacking arbitrary nodes would require most nodes to become faulty before the Byzantine Fault Tolerance threshold would break. If the adversary instead chooses to target the inner consensus group, the threshold is reduced to only 33\% of the consensus group itself, and as such, should be significantly more feasible. Furthermore, hybrid style consensus algorithms are vulnerable to BGP attacks. Combining Nakamoto and BFT style consensus mechanisms, Hybrid style consensus algorithms inherit both groups' vulnerabilities related to BGP.

\textbf{Crash tolerant style} consensus algorithm suffers from the same DDOS susceptibility as BFT style consensus algorithms. The consensus algorithm needs to maintain a stable leader to provide liveness, and such, continuous DDOS attacks targeting leaders can hamper transaction throughput. This may be further exasperated by nodes having limited network throughput. Similar to BFT style consensus algorithms, a DDOS attack may also be used against arbitrary network nodes. In Raft's case, to break the Crash Tolerance threshold, a total of 50\% of the network nodes must become faulty. Thus, unless the network is volatile and at some points contains a lot of faulty nodes, such an attack would be less practical than targeting the leader directly. Finally, Raft is also vulnerable to BGP attacks. An adversary initiating a routing attack against the network leader would be able to completely halt transaction throughput, hindering any broadcasts from reaching the follower nodes. Furthermore, given sufficient power, the adversary could partition the network, attempting to break the Crash Tolerance threshold of 50\%, enabling double spends.

\section{Discussions}

\subsection{Comparison with related work}

Numerous surveys, e.g., \citep{RN21, RN18, RN22, RN17} have been dedicated to present and classify consensus algorithms according to their performance characteristics. In particular, \citep{RN21} and \citep{RN18} present novel evaluation frameworks for analyzing consensus algorithms. \citep{RN21} furthermore highlights the impact network synchrony has on consensus algorithms. Our study differs from related work because it focuses on consensus algorithms' performance characteristics in the specific application scenario in which the network is constrained by varying network throughput. Furthermore, our classification scheme is skewed towards the impact of communication costs and timing assumptions on the algorithms' performance, utilizing the metrics of communication latency, communication complexity, and timing assumptions.

Studies \cite{RN17, RN30, RN21} focuses on the security properties of consensus algorithms, including that of ensuring the integrity of the blockchain. \citep{RN17} discusses core security properties of consensus algorithms and their ability to ensure consistency and tamper-resistance across a blockchain system's ledgers. \citep{RN30} furthermore discusses the circumstances in which these properties may be broken. \citep{RN21} discusses how consensus algorithms may function in a fully asynchronous setting. Our study differentiates from \cite{RN17, RN30, RN21} in that it investigates how consensus algorithms ensure integrity in the specific application scenario in which the network is constrained by varying network throughput. The scenario includes both increased asynchrony of the network, as well as isolation of nodes. It thus provides a unique perspective into which consensus algorithms can provide safety and integrity in such a constrained environment.


\subsection{Known limitations}

This study does not provide insight into experimental data for the different consensus algorithms surveyed. It is, therefore, difficult to draw comparisons regarding transaction throughput among different categories of consensus algorithms.

Furthermore, this study focuses on consensus in enterprise and permissioned blockchain systems affected by limited network throughput. Blockchain systems that do not fit a similar application scenario may emphasize other factors rather than those presented by our classification scheme. The security aspect of this study is based on the reviews of \citep{RN17} and \citep{RN30}. While it is deemed that these works constitute a comprehensive review of security attacks generally applicable to blockchain systems, there may be other security attacks to consider, given the specifics of the blockchain system in question or the scenario in which it is utilized.

\section{Conclusion and Future Work}

The purpose of this study is to analyze how the state of the art consensus algorithms relate to SCM blockchain systems in an enterprise setting with multiple stakeholders, affected by limited network throughput and poor network quality.  The mapping study's classification emphasizes how well the consensus algorithms can satisfy performance, integrity, and security requirements in a degraded network environment. The study gives insights into the advantages and weaknesses of the investigated consensus algorithms. Our future work will evaluate and compare the consensus algorithms identified from this study more quantitatively by running simulations and experiments.

\AtNextBibliography{\footnotesize}
\printbibliography

\end{document}